\documentstyle[12pt]{article}
\begin{document}
\tolerance=5000
\def\be{\begin{equation}}
\def\ee{\end{equation}}
\def\bea{\begin{eqnarray}}
\def\eea{\end{eqnarray}}
\def\nn{\nonumber \\}

\  \hfill
\begin{minipage}{3.5cm}
OCHA-PP-167 \\
NDA-FP-86 \\
November 2000 \\
\end{minipage}

\vfill

\begin{center}
{\large\bf Holographic renormalization group and 
conformal anomaly for AdS$_9$/CFT$_8$ correspondence}

\vfill

{\sc Shin'ichi NOJIRI}\footnote{nojiri@cc.nda.ac.jp},
{\sc Sergei D. ODINTSOV}$^{\spadesuit}$\footnote{
On leave from Tomsk State Pedagogical University, 
634041 Tomsk, RUSSIA. \\
odintsov@ifug5.ugto.mx}, \\
and {\sc Sachiko OGUSHI}$^{\diamondsuit}$\footnote{
JSPS Research Fellow,
g9970503@edu.cc.ocha.ac.jp}
\\

\vfill

{\sl Department of Applied Physics \\
National Defence Academy,
Hashirimizu Yokosuka 239-8686, JAPAN}

\vfill

{\sl $\spadesuit$
Instituto de Fisica de la Universidad de Guanajuato,
Lomas del Bosque 103, Apdo. Postal E-143, 
37150 Leon,Gto., MEXICO }

\vfill

{\sl $\diamondsuit$ Department of Physics,
Ochanomizu University \\
Otsuka, Bunkyou-ku Tokyo 112-0012, JAPAN}

\vfill

{\bf ABSTRACT}

\end{center}

Holographic Renormalization Group (RG) in nine dimensions is considered.
The d8 holographic conformal anomaly is found. It should correspond to
d8 CFT in AdS$_9$/CFT$_8$ correspondence.
The comparison of holographic and QFT anomalies in d8 de Sitter space 
is done. It may give the indication for rigorous AdS$_9$/CFT$_8$ 
correspondence proposal.

\newpage

The Hamilton-Jacobi formalism in higher dimensions provides 
the interesting formulation of 
holographic RG  
\cite{BVV} (for related works, see\cite{related,SOO}).
 This is extremely useful in the study of RG flows\cite{2}
(and references therein) from SG side in AdS/CFT correspondence\cite{AdS}.
In particulary, such investigation gives the possibility to use 
the solutions of multi-dimensional (gauged) supergravity in order to
describe RG flows in dual CFT living on the boundary of corresponding 
AdS spacetime. One of the quantities playing the important 
role in AdS/CFT correspondence \cite{AdS} is holographic conformal
anomaly \cite{ca}.
For example, the comparison of holographic and QFT conformal 
anomalies helps in the explicit identification of
boundary CFT\footnote{The brane quantum gravity presumbly
may be included into this identification as next-to-leading effect \cite{no1}.}
 with the corresponding SG dual. Moreover, the holographic 
 RG is the only known way to get the non-perturbative 
conformal anomaly (which is actually impossible to do in frames of 
usual perturbative QFT). The maximally SUSY Yang-Mills theory is the 
only exclusion from this rule: here QFT calculation 
for exact, non-perturbative result 
is possible.

It is known quite a lot about AdS/CFT correspondence in dimensions 
below 8, say, about AdS$_7$/CFT$_6$, AdS$_5$/CFT$_4$ 
or AdS$_3$/CFT$_2$ set-up (see review
\cite{review} and refs. therein). The corresponding exact CFTs in dimensions 
2,4,6 are explicitly constructed. It would be really unnatural to
expect that Nature is not symmetric and that ,say, 
AdS$_9$/CFT$_8$ correspondence
 and corresponding exact CFT do not occur. Hence, 
it would be of great interest to
extend the corresponding results to higher dimensions: d9, d11 (where 
M-theory is presumbly residing)\footnote{Despite some attempts d8 CFT 
 is not constructed yet explicitly.
The multiplet of CFT$_8$ could be one including rank 3 (with 3 vector 
indeces and maybe anti-symmetric) tensor field. We need to 
consider what is the low energy effective theory in D7-brane. 
Let us consider the situation where $N$ branes are piled up. 
If a (open) string connect two branes, the string corresponds to 
vector field in the low energy and the field contains $N^2$ 
components, which correspond to ${\cal N}=4$ Yang-Mills theory 
in 4 dimensions. If strings with junction (3 strings with 
one vertex) connect 3 branes, it corresponds to rank 2
tensor field with $N^3$ components, which lead to CFT$_6$. 
If strings (or maybe strings with two junctions or verteces) connect 4 
branes, it would correspond 
to rank 3 tensor field with $N^4$ components, which would 
correspond to CFT$_8$. Note that the number of the 
vector indeces corresponds to the directions from one fixed brane 
to the other branes, then the number of the indeces is the number 
of the connected branes minus one. Another way of doing things could be
 to start from d10 supergravity, to compactify it to 8 dimensions and to try
to construct something like modified 8d vector multiplet. Clearly, it is not 
an easy task.}. As one step in this direction 
one can calculate the holographic conformal anomaly in higher dimensions. 
 In the present Letter, starting from the systematic
prescription for solving the Hamilton-Jacobi equation
(i.e. flow equation) given in \cite{FMS}, we will perform 
such a calculation in eight dimensions.\footnote{The very interesting 
attempt based on counterterm method to calculate d8 conformal anomaly 
has been performed in ref.\cite{BV}. However, the explicit result was not
obtained there. Moreover, it is known \cite{SOO} that counterterm method 
in higher dimensions 
may lead to some ambigious result.}

First, we briefly review the formulation discussed in
\cite{BVV, FMS}.  
 One starts from $d+1$ dimensional AdS-like metric in the following
form 
\be
\label{met}
ds^{2} = G_{MN}dX^{M}dX^{N} = 
dr^{2} +G_{\mu\nu}(x,r)dx^{\mu}dx^{\nu} .
\ee
where $X^{M}=(x^{\mu},r)$ with $\mu,\nu=1,2,\cdots ,d$.
The action on a $(d+1)$ dimensional manifold $M_{d+1}$ with the
boundary $\Sigma_{d}=\partial M_{d+1}$ is given by
\bea
S_{d+1}&=& \int_{M_{d+1}} d^{d+1}x\sqrt{G}(V-R) -
2\int_{\Sigma_{d}} d^{d}x \sqrt{G}K \nn
&=& \int_{\Sigma_{d}}d^{d}x \int dr \sqrt{G}\left( 
V-R+K_{\mu\nu}K^{\mu\nu}-K^{2} \right) \nn
&\equiv & \int d^{d}x dr \sqrt{G} {\cal L}_{d+1}.
\eea
where $R$ and $K_{\mu\nu}$ are the scalar curvature and 
the extrinsic curvature on $\Sigma_{d}$ respectively.
$K_{\mu\nu}$ is given as
\bea
K_{\mu\nu}={1\over 2}{\partial G_{\mu\nu} \over \partial r},\quad
K=G^{\mu\nu}K_{\mu\nu}
\eea
In the canonical formalism, ${\cal L}_{d+1}$ is rewritten
by using the canonical momenta $\Pi_{\mu\nu}$
and Hamiltonian density ${\cal H}$ as
\be
{\cal L}_{d+1} = \Pi^{\mu\nu}{\partial G_{\mu\nu} \over \partial r}
+{\cal H} \ ,\quad
{\cal H} \equiv {1 \over d-1}(\Pi ^{\mu}_{\mu})^2-\Pi_{\mu\nu}^{2}
+V-R\ . 
\ee
The equation of motion for $\Pi ^{\mu\nu}$ leads to 
\bea
\Pi^{\mu\nu}=K^{\mu\nu}-G^{\mu\nu}K .
\eea
The Hamilton constraint ${\cal H}=0$ leads to the
Hamilton-Jacobi equation (flow equation) 
\bea
\label{HJ}
\{ S,S \} (x) &=& \sqrt{G} {\cal L}_{d} (x) \\
\{ S,S \} (x) &\equiv & {1 \over \sqrt{G}}
\left[-{1 \over d-1}\left(G_{\mu\nu}{\delta S \over \delta G_{\mu\nu}}
 \right)^{2}+\left( {\delta S \over \delta G_{\mu\nu}}
 \right)^{2} \right] , \\
{\cal L}_{d}(x) &\equiv & V -R[G].
\eea

One can decompose the action $S$ into a local and non-local part
discussed in ref.\cite{BVV} as follows
\bea
S[G(x)] &=& S_{loc}[G(x)]+\Gamma[G(x)] ,
\eea
 Here $S_{loc}[G(x)]$ is tree level action and $\Gamma$ contains
the higher-derivative and non-local terms. 
In the following discussion, we take the systematic method 
of ref.\cite{FMS}, which is weight calculation.
  The $S_{loc}[G]$ can be expressed as a sum of local terms
\be
S_{loc}[G(x)] = \int d^{d}x \sqrt{G} {\cal L}_{loc}(x) 
= \int d^{d}x \sqrt{G} 
\sum _{w=0,2,4,\cdots} [{\cal L}_{loc}(x)]_{w} 
\ee
The weight $w$ is defined by following rules;
\[
G_{\mu\nu },\; \Gamma : \mbox{weight 0} \ ,\quad
\partial_{\mu} : \mbox{weight 1} \ ,\quad
R,\; R_{\mu\nu} : \mbox{weight 2} \ ,\quad
{\delta \Gamma \over \delta G_{\mu\nu}} : 
\mbox{weight $d$} \ .
\]
Using these rules and (\ref{HJ}),  one obtains 
the equations, which depend on the weight as
\bea
\label{wt1}
\sqrt{G}{\cal L}_{d} &=& \left[ \{ S_{loc},S_{loc} \} \right]_0 +
\left[ \{ S_{loc},S_{loc} \} \right]_2 \\
\label{wt2}
0 &=& \left[ \{ S_{loc},S_{loc} \} \right]_w  
\quad (w=4,6,\cdots d-2), \\
\label{wt3}
0 &=& 2\left[ \{ S_{loc}, \Gamma \} \right]_d 
+ \left[ \{ S_{loc},S_{loc} \} \right]_d
\eea
The above equations which determine $\left[ {\cal L}_{loc} \right]_{w}$. 
$\left[ {\cal L}_{loc} \right]_{0}$ and $[{\cal L}_{loc}]_{2}$ are
parametrized by 
\be
 \left[ {\cal L} _{loc} \right]_0 = W\ ,\quad 
 \left[ {\cal L} _{loc} \right]_2 = -\Phi R \ .
\ee
Thus one can solve (\ref{wt1}) as
\bea
V = -{d \over 4(d-1)}W^{2} \ ,\quad 
-1 = {d-2 \over 2(d-1)} W\Phi\ .
\eea
Setting $V=2\Lambda =-d(d-1)/l^{2}$, where 
$\Lambda$ is the bulk cosmological constant 
and the parameter $l$ is the radius of 
the asymptotic AdS$_{d+1}$, we obtain $W$ and $\Phi$ as
\bea
W=-{2(d-1) \over l},\quad \Phi = {l \over d-2}.
\eea
To obtain the higher weight ($w \ge 4$) local terms 
related with conformal anomaly,
we introduce a local term $\left[ {\cal L}_{loc} \right]_{4}$
as follows
\bea
\left[ {\cal L}_{loc} \right]_{4}=X R^2 +Y R_{\mu \nu }R^{\mu \nu}
+Z R_{\mu\nu\lambda \sigma }R^{\mu\nu\lambda \sigma }.
\eea
Here $X,Y$ and $Z$ are some constants determined 
by (\ref{wt2}).  
The calculation of $\left[\{ S_{loc},S_{loc} \} \right]_{4}$
was done in \cite{FMS} as 
\bea
\label{ano4}
\lefteqn{{1 \over \sqrt{G}}\left[\{ S_{loc},S_{loc} \} \right]_{4} 
=-{W \over 2(d-1)}\left( (d-4)X-{dl^{3} \over 4(d-1)(d-2)^{2}}
\right) R^{2} } \nn
&&-{W \over 2(d-1)}\left( (d-4)Y +{l^{3} \over (d-2)^{2}}
\right)R_{\mu\nu}R^{\mu\nu}-{d-4 \over 2(d-1)}WZR_{\mu\nu\lambda\sigma}
R^{\mu\nu\lambda\sigma} \nn
&&+\left( 2X +{d \over 2(d-1)}Y+{2\over d-1}Z \right) . 
\eea
For $d \ge 6$, from $\left[\{ S_{loc},S_{loc} \} \right]_{4}=0$
one finds 
\bea
\label{XYZ}
X={d l^3 \over 4(d-1)(d-2)^{2}(d-4) }, 
\quad Y=-{l^3 \over (d-2)^{2} (d-4)}, 
\quad Z=0.
\eea
Using them, one can calculate 
$\left[\{ S_{loc},S_{loc} \} \right]_{6}$ as \cite{FMS} 
\bea
\label{ano6}
\lefteqn{{1 \over \sqrt{G}}\left[\{ S_{loc},S_{loc} \} \right]_{6} 
=\Phi \left[ \left(-4X +{d+2 \over 2(d-1)}Y \right) 
RR_{\mu\nu}R^{\mu\nu} +{d+2 \over 2(d-1)}XR^{3} \right.} \nn
&& -4YR^{\mu\lambda}R^{\nu\sigma}R_{\mu\nu\lambda\sigma} 
 +(4X+2Y)R^{\mu\nu}\nabla_{\mu}\nabla_{\nu}R-2YR^{\mu\nu}
\nabla^{2}R_{\mu\nu} \nn
&& \left.+\left(-2X-{d-2 \over 2(d-1)}Y \right)
R\nabla^{2}R \right] 
+(\mbox{contributions from }[{\cal L}_{loc}]_{6} ) \nn
&=&l^{4}\left[-{3d+2 \over 2(d-1)(d-2)^{2}(d-4)}RR_{\mu\nu}R^{\mu\nu} 
+{d(d+2) \over 8(d-1)^{2}(d-2)^{3}(d-4)}R^{3} \right. \nn 
&&+{4\over (d-2)^{3}(d-4)}R^{\mu\lambda}R^{\nu\sigma}
R_{\mu\nu\lambda\sigma}-{1\over (d-1)(d-2)^{2}(d-4)}
R^{\mu\nu}\nabla_{\mu}\nabla_{\nu}R \nn
&& \left.+{2 \over (d-2)^{3}(d-4)}R^{\mu\nu}\nabla^{2}R_{\mu\nu}
-{1 \over (d-1)(d-2)^{3}(d-4)}R\nabla^{2}R \right] \nn
&&+(\mbox{contributions from }[{\cal L}_{loc}]_{6} ) .
\eea
The weight $d$ flow equation (\ref{wt3}), which is related 
with the conformal anomaly in $d$ dimensions \cite{BVV,FMS}, is
written by
\bea
\label{ano1}
-{W \over 2(d-1)}{1\over \sqrt{G}}
G_{\mu\nu}{\delta \Gamma \over \delta G_{\mu\nu}}
=-\left[ \{ S_{loc} , S_{loc} \} \right]_{d} .
\eea 
This $G_{\mu\nu}{\delta \Gamma \over \delta G_{\mu\nu}}$
can be regarded as the sum of conformal anomaly ${\cal W}_{d}$ 
and the total derivative term $\nabla_{\mu}{\cal J}^{\mu}_{d}$
in $d$ dimensions.  Thus we rewrite (\ref{ano1}) as following
\bea
\kappa^2{\cal W}_{d}+\nabla_{\mu}{\cal J}^{\mu}_{d}={d-1 \over W \sqrt{G}}
\left[ \{S_{loc},S_{loc} \}\right]_{d} .
\eea
Here $\kappa^2$ is $d+1$ dimensional gravitational coupling.
Using the above relation,  one can get the holographic conformal 
anomaly in
$4$ dimensions from (\ref{ano4}):
\be
\kappa^2{\cal W}_{4} = -{l\over 2\sqrt{G} }
\left[ \{S_{loc},S_{loc} \}\right]_{4} 
= l^{3}\left( {1\over 24}R^{2} -{1\over 8} R_{\mu\nu}R^{\mu\nu}
\right)\ .
\ee
This agrees with the result in Refs.\cite{ca} 
calculated  by various methods (using
AdS/CFT duality).  Further, the above calculation can be
extended to include dilaton (a scalar). 
Note that holographic conformal anomaly for the presence of dilaton
turns out to be scheme-dependent \cite{SOO}. 
The conformal anomaly in $6$ dimensions is
calculated from (\ref{ano6}) as
\bea
\kappa^2{\cal W}_{6} &=& -{l\over 2\sqrt{G} }
\left[ \{S_{loc},S_{loc} \}\right]_{6} \nn 
&=& l^{5} \left( {1\over 128}RR_{\mu\nu}R^{\mu\nu}-{3\over 3200}R^{3}
-{1\over 64}R^{\mu\lambda}R^{\nu\sigma}R_{\mu\nu\lambda\sigma} \right.\nn
&&+ \left. {1\over 320}R^{\mu\nu}\nabla_{\mu}\nabla_{\nu}R
-{1\over 128}R^{\mu\nu}\nabla^{2}R_{\mu\nu}
+{1\over 1280}R\nabla^{2}R  \right),
\eea
which coincides exactly with $6$ dimensional conformal anomaly
in \cite{ca}. Above discussions have already been performed 
in Ref.\cite{FMS}.

The local terms of weight $6$: $\left[ {\cal L}_{loc} \right]_{6}$
is assumed to be
\bea
\lefteqn{\left[ {\cal L}_{loc} \right]_{6}=a R^3 +b R R_{\mu \nu }R^{\mu \nu}
+c R R_{\mu\nu\lambda \sigma }R^{\mu\nu\lambda \sigma }+e 
R_{\mu\nu\lambda \sigma}R^{\mu\rho}R^{\nu \sigma}} \\
&&+f \nabla_{\mu}R\nabla^{\mu}R
+g \nabla_{\mu}R_{\nu\rho}\nabla^{\mu}R^{\nu\rho}
+h \nabla_{\mu}R_{\nu\rho\sigma\tau}\nabla^{\mu}R^{\nu\rho\sigma\tau}
+j R^{\mu\nu}R^{\rho}_{\nu}R_{\rho\mu}.\nonumber
\eea
Adding above terms to (\ref{ano6}), we obtain 
\bea
\lefteqn{{1 \over \sqrt{G}}\left[\{ S_{loc},S_{loc} \} \right]_{6} 
= \left( b\left( {d\over 2}-3 \right){2\over l}
-{(3d+2)l^{4} \over 2(d-1)(d-2)^{3}(d-4)}
\right) RR_{\mu\nu}R^{\mu\nu}} \nn
&&+\left( a\left( {d\over 2}-3 \right){2\over l}
+{d(d+2)l^{4} \over 8(d-1)^{2}(d-2)^{3}(d-4) }\right)R^{3} \nn
&& +\left( \left\{ -e\left( {d\over 2}+2 \right)
 -2g -{3j\over 2}(2-d)
\right\}{2\over l} +{4l^{4} \over (d-2)^{3}(d-4)} \right)
R^{\mu\lambda}R^{\nu\sigma}R_{\mu\nu\lambda\sigma} \nn
&& +\left( \left\{ b(2-d) -4c+e\left( {d\over 2}-1 \right)
+{3j\over 2}(2-d)\right\} {2\over l} \right.\nn
&& \left. -{l^{4} \over (d-1)(d-2)^{2}(d-4) } \right)
R^{\mu\nu}\nabla_{\mu}\nabla_{\nu}R \nn
&& +\left( \left\{ 2b(1-d)-de+2g-3j\right\} {2\over l}
+{2l^4 \over (d-2)^{3}(d-4)}\right)
R^{\mu\nu}\nabla^{2}R_{\mu\nu}\nn
&&+\left(\left\{ 6a(1-d)-b\left( 1+{d\over 2} \right) 
 -2c-{1\over 2}e+2f  \right\} {2\over l} \right.\nn
&&\left. -{l^{4}\over (d-1)(d-2)^{3}(d-4) } \right)R\nabla^{2}R \nn
&& +\left( {d\over 2}g+2h+2f(d-1)  \right){2\over l} \nabla^{4}R 
+\left( {d \over 2} -3\right){2c \over l}
RR_{\mu\nu\rho\sigma}R^{\mu\nu\rho\sigma} \nn
&& +\left( 6a(1-d)-db -4c -{3\over 4}e+\left( {d\over 2}-1 \right)f 
 -{g\over 2}+{3\over 8}(2-d)j  \right)
{2\over l}\nabla_{\mu}R\nabla^{\mu}R \nn
&& +\left( 2b(1-d)+2e(1-d) +g\left( {d\over 2}-1 \right)
 -8h-3j \right){2\over l}
\nabla_{\kappa}R_{\mu\nu}\nabla^{\kappa}R^{\mu\nu} \nn
&& +\left( (2d-3)e+2g+8h +{3\over 2}(2-d)j \right){2\over l}
\nabla_{\kappa}R^{\mu\nu}\nabla_{\nu}R_{\mu}^{\kappa} \nn 
&& +\left( (d-1)e+2g-dj \right)
{2\over l}R^{\mu\nu}R_{\nu}^{\rho}R_{\mu\rho} 
+(2-d){2e\over l}
R^{\mu\nu\rho\sigma} \nabla _{\mu}\nabla_{\rho}R_{\nu\sigma} \nn
&& + \left( 2c(1-d)+\left({d\over 2} -1 \right)h \right)
{2 \over l} \nabla_{\alpha} R_{\mu\nu\rho\sigma}
\nabla^{\alpha}R^{\mu\nu\rho\sigma} \nn
&& +\left( 2c(1-d)+2h \right){2 \over l}
R_{\mu\nu\rho\sigma} \nabla^{2}R^{\mu\nu\rho\sigma} \nn
&& +\left( 4R^{\mu\rho\sigma\tau} R_{\lambda\mu} 
R^{\lambda}_{\rho\sigma\tau} 
 -4 R^{\mu\rho\sigma\tau} R^{\nu}_{\mu\lambda\rho} 
R^{\lambda}_{\nu\tau\sigma} \right){2h \over l} \nn
&& +\left( -8 R^{\mu\rho\sigma\tau} R^{\nu}_{\mu\lambda\sigma} 
R^{\lambda}_{\tau\nu\rho}+4 \nabla_{\nu} R_{\mu\rho\sigma\tau} 
\nabla^{\mu}R^{\nu\rho\sigma\tau}\right){2h \over l} \nn
\lefteqn{= \left[ \left( {(d-6)b \over l}
 -{(3d+2) l^{4} \over 2(d-1)(d-2)^{2}(d-4)}
\right) RR_{\mu\nu}R^{\mu\nu} \right.} \nn
&& +\left( {(d-6)a \over l} +{d(d+2) l^{4} \over 
8(d-1)^{2}(d-2)^{3}(d-4)}\right)R^{3} 
+\left({(d-6)c \over l } \right)RR_{\mu\nu\lambda \sigma }
R^{\mu\nu\lambda \sigma } \nn 
&& +\left({(d-6)e \over l} + {16h \over l}
+{4l^{4} \over (d-2)^{3}(d-4)}\right) R^{\mu\lambda}R^{\nu\sigma}
R_{\mu\nu\lambda\sigma} \nn
&&+\left(-\left(3 - {d \over 2}\right){2f \over l} 
+ {4h \over l} + {d l^{4} \over 2(d-1)(d-2)^{3}(d-4)}\right)
\nabla^\mu R \nabla_{\mu}R \nn
&& + \left({(d-6)g \over l} - {2 l^{4}\over (d-2)^{3}(d-4)} 
 + {16 h \over l} \right)
 \nabla^\rho R^{\mu\nu}\nabla_\rho R_{\mu\nu} \nn
&& +\left({2j \over l}\left( {d\over 2}-3 \right)
 - {16h \over l}\right) R^{\mu\nu}R^{\rho}_{\nu}R_{\rho\mu}\nn 
&& \left. +\left({(6-d) h \over l} \right)
R_{\mu\nu\lambda \sigma }\nabla^{2} R^{\mu\nu\lambda \sigma }
\right] + \mbox{total derivative terms}\ .
\eea
For $d \ge 8$, from $\left[\{ S_{loc},S_{loc} \} \right]_{6}=0$,
if one neglects the total derivative terms, 
 the coefficients $a,b,c,e,f,g,h,j$ are
\bea
\label{abc}
&& a= -{d(d+2) l^{5} \over 8(d-1)^{2}(d-2)^{3}(d-4)(d-6)} \nn
&& b= {(3d+2) l^{5} \over 2(d-1)(d-2)^{2}(d-4)(d-6)} \ ,\quad 
c= 0 \nn
&& e= -{4 l^{5} \over (d-2)^{3}(d-4)(d-6)} \ ,\quad
f= -{d l^{5} \over 2(d-1)(d-2)^{3}(d-4)(d-6)} \nn
&& g= {2 l^{5} \over (d-2)^{3}(d-4)(d-6) } \ ,\quad 
h= 0\ ,\quad j= 0.
\eea
We can also  consider $d=8$ case in the same way.
In $d=8$ case,  
one obtains ${1\over \sqrt{G}}[\{ S_{loc},S_{loc} \}]_{8}$
as follows
\bea
\lefteqn{{1\over \sqrt{G}}[\{ S_{loc},S_{loc} \}]_{8} 
= \left( -{(d+8)X^{2} \over 4(d-1)}+ {(d+4)al 
\over 2(d-1)(d-2)} \right) R^{4} } \nn
&& + \left( 2X^2  +{(-d+4) XY \over 2(d-1) }
 -{6al \over (d-2)} 
+{(4-d)bl \over 2(d-1)(d-2)} +{el\over 2(d-1)(d-2)}
\right.\nn
&& \left. -{2fl \over (d-1)(d-2)} \right) R^{2}\nabla^{2} R 
+ \left(-{(d+8) Y^{2} \over 4(d-1)}-{2bl \over d-2 } \right)
(R^{\mu\nu}R_{\mu\nu})^{2} \nn
&& + \left(-4X^2 -{d \over 4(d-1) }Y^2 
-2 X Y \right)(\nabla^{2} R)^{2} \nn 
&& +\left( 4X^2 -{(d + 8 ) XY \over 2(d-1)} -{6al \over (d-2)}
+{(d+4)bl \over 2(d-1)(d-2)} \right)R^{2}R^{\mu\nu}R_{\mu\nu} \nn
&& + \left( 4X^2 +Y^2  +4 XY \right) 
\nabla^{\mu}\nabla^{\nu}R\nabla_{\mu}\nabla_{\nu}R \nn
&&+\left( -8X^2  -4XY +{12al \over d-2 }+ {bl\over d-1}
+{del \over 2(d-1)(d-2)}
\right) R R^{\mu \nu}\nabla_{\mu}\nabla_{\nu} R \nn
&&+\left({(d-4)(-2d+1) Y^{2} \over 2(d-1)}+ 2 XY
 -{2bl\over d-2}-{el \over d-2}
+{4fl \over d-2}  \right)R^{\mu\nu}R_{\mu\nu} \nabla^{2}R \nn
&&+\left( -2Y^2 -4XY \right)
\nabla ^{2}R^{\mu\nu}\nabla_{\mu}\nabla_{\nu}R 
+\left( 4Y^2 -{2el\over d-2}+{4gl \over d-2} \right)
\nabla ^{2}R^{\mu\nu}R_{\mu \lambda \nu \kappa}R^{\lambda \kappa} \nn
&&+ Y^{2} \nabla ^{2}R_{\mu\nu}\nabla ^{2}R^{\mu\nu}
+ \left( 4Y^2 +{4el \over d-2} \right) R^{\lambda}_{\mu \kappa \nu}
R^{\mu\nu}R^{\kappa}_{\sigma \lambda \gamma}R^{\sigma \gamma} \nn
&&+ \left(-4Y^2 -8XY \right) R^{\lambda }_{\mu \kappa \nu}R^{\mu \nu}
\nabla_{\lambda}\nabla^{\kappa} R 
+{4el\over d-2}R_{\kappa\lambda} R^{\kappa}_{\sigma\mu\nu} 
\nabla^{\mu} \nabla ^{\lambda} R^{\nu\sigma} \nn
&&+\left( 8XY -{4bl \over d-2}
+{(-d+6)el \over 2(d-1)(d-2)}+{2gl \over (d-1)(d-2)}
\right) R^{\kappa}_{\lambda}R^{\lambda}_{\mu \kappa \nu}R^{\mu \nu}R \nn
&&+\left(4 XY -{4bl \over (d-2)}-{el\over (d-1)}
 -{2gl\over (d-1)(d-2)} \right) R_{\kappa \lambda} R\nabla^{2} 
R^{\kappa \lambda}\nn
&&+\left( -{6al\over d-2}-{bl\over d-1}+{3el\over 4(d-1)(d-2)}
+{dfl\over 2(d-1)(d-2)}\right. \nn
&&\left.+{gl\over 2(d-1)(d-2)} \right) R (\nabla R )^{2} 
+{l \over d-2}\left( 4b + 2e + 2g \right)
R_{\kappa\lambda} R^{\mu\kappa} \nabla_{\mu}\nabla^{\lambda} R \nn
&&+{l \over d-2}\left(12a + 2b + {e \over 2}
 - 2f \right) R_{\mu\nu}\nabla ^{\mu}R \nabla^{\nu} R \nn
&& + {l \over d-2}\left( -2b - 2e + {dg \over 2(d-1)}
\right) R \nabla _{\kappa}R_{\mu\nu} 
\nabla^{\kappa} R^{\mu\nu} 
+ \left( {2fl\over d-2}+ {gl\over 2(d-1)} \right) R \nabla^{4} R \nn
&& - {( 4f + 2g )l \over d-2} 
R_{\mu\nu} \nabla^{\mu}\nabla^{\nu}\nabla^{2} R 
+ {( 4b + 2e )l \over d-2}
R_{\kappa\lambda}R^{\mu\nu}
\nabla^{\kappa}\nabla^{\lambda}R_{\mu\nu} \nn
&& + {l \over d-2}\left( 4b + 4e - 2g \right)
R_{\kappa\lambda}\nabla^{\kappa}R_{\mu\nu}\nabla^{\lambda}R^{\mu\nu} 
+ {l \over d-2}\left( 4b - 2g \right)
R_{\kappa\lambda}\nabla_{\mu}R \nabla^{\lambda}R^{\mu\kappa} \nn
&& - {l \over d-2}\left( 4b + 2e \right)
R_{\kappa\lambda}\nabla_{\mu}R \nabla^{\mu}R^{\kappa\lambda } 
+ {l \over d-2}\left( e - {2g \over d-1} \right)
R R^{\mu\nu}R_{\kappa\mu}R^{\kappa \nu} \nn
&& + {l \over (d-1)(d-2)}\left( (2d-1) e - 2g \right)
R \nabla^{\mu}R^{\nu \kappa} \nabla_{\nu} R_{\mu\kappa} \nn
&&-{del \over (d-1)(d-2)}
RR^{\mu\nu \kappa\lambda}\nabla_{\mu} \nabla_{\kappa}R_{\nu\lambda} 
+{2el \over d-2} 
R_{\mu\nu}\nabla_{\kappa} R^{\nu\rho }\nabla_{\rho}R^{\kappa\mu} \nn
&& -{4(e+g)l \over d-2}
R_{\kappa\lambda} R^{\lambda}_{\nu\rho\mu}R^{\kappa\rho}R^{\nu \mu} 
 - {8gl \over d-2} R_{\kappa\lambda} 
\nabla_{\mu}R^{\lambda \nu} \nabla^{\mu} R_{\nu}^{\kappa}
 - {4gl \over d-2}
R_{\kappa\lambda} R_{\nu}^{\kappa} \nabla^{2} R^{\lambda \nu} \nn
&& -{4(e+g)l \over d-2} 
R_{\kappa\lambda}R_{\nu}^{\mu} \nabla_{\mu} \nabla^{\lambda} 
R^{\kappa \nu} -{4(e-g)l \over d-2} R_{\kappa\lambda} 
\nabla_{\mu}R^{\kappa \nu} \nabla^{\lambda} R_{\nu}^{\mu} \nn
&&  -{(2e - 4g)l \over d-2} R_{\kappa}^{\lambda} 
R^{\mu \nu} \nabla^{2} R^{\kappa}_{\mu \lambda \nu} 
+{2gl\over d-2}R_{\mu\nu}\nabla^{4}R^{\mu\nu} \nn
&& +{4el \over d-2}
R_{\kappa\lambda} \nabla^{\mu}R^{\nu \sigma} \nabla^{\lambda} 
R^{\kappa}_{\sigma \mu \nu} 
 -{(4e-8g)l \over d-2}
R^{\kappa\lambda} \nabla^{\mu}R^{\nu}_{\sigma} \nabla_{\mu} 
R^{\sigma}_{\kappa \nu \lambda} \\
&&+{(2e+4g)l \over d-2}
R_{\kappa \lambda}R^{\kappa}_{\rho}R^{\rho}_{\nu}R^{\nu\lambda}
+{4el \over d-2}
\left( R_{\kappa \lambda}R^{\nu}_{\sigma}R^{\rho\sigma\lambda\mu}
R^{\kappa}_{\rho\mu\nu}+
R_{\kappa \lambda}R_{\nu}^{\sigma}R^{\rho \nu \lambda \mu}
R^{\kappa}_{\sigma\mu\rho}\right)\ .\nonumber
\eea
Substituting $X,Y,a,b,e,f,g,j$ in (\ref{XYZ}) and (\ref{abc})
into the above equation and putting $d=8$, we obtain the explicit form of 
$\left[\{ S_{loc},S_{loc} \} \right]_{8}$ 
and conformal anomaly in $8$ dimensions 
\bea
\label{8dan}
\lefteqn{-{2 \over l^7}\kappa^2{\cal W}_{8} 
 ={1\over l^6\sqrt{G}}[\{ S_{loc},S_{loc} \}]_{8} }\nn
&& = {13 \over 889056}  R^{4}
 - {1 \over 2352 } R^{2}\nabla^{2} R 
 - {79 \over 36288} (R^{\mu\nu}R_{\mu\nu})^{2}
 - {1 \over 508032 } (\nabla^{2} R)^{2} \nn
&& + {53 \over 63504} R^{2}R^{\mu\nu}R_{\mu\nu} 
+ { 1 \over 112896} \nabla^{\mu}\nabla^{\nu}R
\nabla_{\mu}\nabla_{\nu}R 
+ {61 \over 63504 } R R^{\mu \nu}\nabla_{\mu}\nabla_{\nu} R \nn
&& - {23 \over 10368} R^{\mu\nu}R_{\mu\nu} \nabla^{2}R 
 -{1 \over 24192} \nabla ^{2}R^{\mu\nu}\nabla_{\mu}\nabla_{\nu}R 
+ {1 \over 576} \nabla ^{2}R^{\mu\nu}R_{\mu \lambda \nu \kappa}
R^{\lambda \kappa} \nn
&& + {1 \over 20736}\nabla ^{2}R_{\mu\nu}\nabla ^{2}R^{\mu\nu} 
 -  {7 \over 5184} R^{\lambda}_{\mu \kappa \nu}
R^{\mu\nu}R^{\kappa}_{\sigma \lambda \gamma}R^{\sigma \gamma} 
 - {1 \over 12096} R^{\lambda }_{\mu \kappa \nu}R^{\mu \nu}
\nabla_{\lambda}\nabla^{\kappa} R \nn
&& -{1 \over 648 }R_{\kappa\lambda} R^{\kappa}_{\sigma\mu\nu} 
\nabla^{\mu} \nabla ^{\lambda} R^{\nu\sigma} 
 -{13 \over 3024} R^{\kappa}_{\lambda}R^{\lambda}_{\mu \kappa \nu}
 R^{\mu \nu}R  - {37 \over 9072} R_{\kappa \lambda} R\nabla^{2} 
 R^{\kappa \lambda} \nn
&& -{31 \over 28224} R (\nabla R )^{2} 
+{71 \over 18144} R_{\kappa\lambda} R^{\mu\kappa} 
\nabla_{\mu}\nabla^{\lambda} R 
+{65 \over 28224} R_{\mu\nu}\nabla ^{\mu}R \nabla^{\nu} R \nn
&& - {23 \over 18144} R \nabla _{\kappa}R_{\mu\nu} 
\nabla^{\kappa} R^{\mu\nu} - {1 \over 72576} R \nabla^{4} R
-{1\over 6048}  R_{\mu\nu} \nabla^{\mu}\nabla^{\nu}\nabla^{2} R \nn
&& + {2\over 567} R_{\kappa\lambda}R^{\mu\nu}
\nabla^{\kappa}\nabla^{\lambda}R_{\mu\nu} + {43 \over 18144} 
R_{\kappa\lambda}\nabla^{\kappa}R_{\mu\nu}\nabla^{\lambda}R^{\mu\nu} 
\nn
&& + {71 \over 18144} 
R_{\kappa\lambda}\nabla_{\mu}R \nabla^{\lambda}R^{\mu\kappa} 
 - {2\over 567} R_{\kappa\lambda}\nabla_{\mu}R \nabla^{\mu}
R^{\kappa\lambda } -{1\over 2268} R R^{\mu\nu}R_{\kappa\mu}
R^{\kappa \nu} \nn
&& -{1\over 1134} R \nabla^{\mu}R^{\nu \kappa} 
\nabla_{\nu} R_{\mu\kappa}  +{1\over 2268}
RR^{\mu\nu \kappa\lambda}\nabla_{\mu} 
\nabla_{\kappa}R_{\nu\lambda} \nn
&& -{1 \over 1296} 
R_{\mu\nu}\nabla_{\kappa} R^{\nu\rho }\nabla_{\rho}R^{\kappa\mu} 
+ {1\over 1296} 
R_{\kappa\lambda} R^{\lambda}_{\nu\rho\mu}R^{\kappa\rho}R^{\nu \mu} \nn
&&-{1\over 648}
R_{\kappa\lambda }\nabla_{\mu}R^{\lambda\nu}\nabla^{\mu}R^{\kappa}_{\nu}
-{1\over 1296} R_{\kappa\lambda} R_{\nu}^{\kappa} \nabla^{2} R^{\lambda \nu}
+{1\over 1296} 
R_{\kappa\lambda}R_{\nu}^{\mu} \nabla_{\mu} \nabla^{\lambda} 
R^{\kappa \nu}\nn
&&+ {1\over 432} 
R_{\kappa\lambda} \nabla_{\mu}R^{\kappa \nu} \nabla^{\lambda} R_{\nu}^{\mu} 
+{1\over 648}+R_{\kappa}^{\lambda} R^{\mu \nu} \nabla^{2} R^{\kappa}_{\mu \lambda \nu}+{1\over 2592}R_{\mu\nu}\nabla^{4}R^{\mu\nu} \nn 
&&-{1 \over 648}
R_{\kappa\lambda} \nabla^{\mu}R^{\nu \sigma} \nabla^{\lambda} 
R^{\kappa}_{\sigma \mu \nu} 
+{1 \over 324} 
R^{\kappa\lambda} \nabla^{\mu}R^{\nu}_{\sigma} \nabla_{\mu} 
R^{\sigma}_{\kappa \nu \lambda} \nn
&&-{1 \over 648}
\left( R_{\kappa \lambda}R^{\nu}_{\sigma}R^{\rho\sigma\lambda\mu}
R^{\kappa}_{\rho\mu\nu}+
R_{\kappa \lambda}R_{\nu}^{\sigma}R^{\rho \nu \lambda \mu}
R^{\kappa}_{\sigma\mu\rho}\right).
\eea
As one can see already in eight dimensions (and omitting total
derivative terms) the explicit result for holographic conformal
anomaly is quite complicated. It is clear that going to higher 
dimensions it is getting much more complicated.

As an example, we consider de Sitter space, where
curvatures are covariantly constant and given by
\be
\label{curvs}
R_{\mu\nu\rho\sigma}={ 1\over l^2}\left(g_{\mu\rho}g_{\nu\sigma} 
 - g_{\mu\sigma}g_{\nu\rho}\right) \ ,\quad
R_{\mu\nu}={d-1 \over l^2}g_{\mu\nu} \ ,\quad 
R={d(d-1) \over l^2}\ .
\ee
Here $l$ is the radius of the de Sitter space\footnote{The choice
of de Sitter space is caused by several reasons. First, the explicit 
expression for holographic d8 CA is significally simplified on
this highly symmetric background. Second, the quantum field calculation 
of d8 conformal anomaly has been done so far only for de Sitter background 
as we mention in the text. Third, even in d4 case the calculation of CA 
on de Sitter space has been done by various authors because it may have 
the interesting applications in cosmology/BH physics and it is related
 with Casimir energy.}
 and it is related to 
the cosmological constant $\Lambda$ by
$\Lambda={(d-2)(d-1) \over l^2}$. 
By putting $d=8$ in (\ref{curvs}) and substituting the  
 curvatures into (\ref{8dan}), we find an expression for the 
anomaly:
\be
\label{8danS}
\kappa^2{\cal W}_{8} =-{l\over 2\sqrt{G}}[\{ S_{loc},S_{loc} \}]_{8}
 = - {62069 \over 1296 l}\ .
\ee
We should note that ${1 \over \kappa^2}$ is 9 dimensional one
here, then $\kappa^2$ 
has the dimension of 7th power of the length. 

In refs.\cite{CT,BEO} the QFT conformal anomalies coming from scalar 
and spinor fields in 8d de Sitter space are found
\be
\label{SSano}
T_{\rm scalar}= -{23 \over 34560 \pi^4 l^8}\ ,\quad 
T_{\rm spinor}= -{2497 \over 34560 \pi^4 l^8}\ .
\ee
If there is supersymmetry, the number of the scalars 
is related with that of the spinors. For example, consider the  
matter supermultiplet and take only scalar-spinor part of it
(one real scalar and one Dirac spinor) as vector is not conformally invariant
in d8 dimensions.
If there is $N^4$ pairs of scalars and spinors\footnote{ In
 AdS$_5$/CFT$_4$ correspondence where exact CFT is maximally SUSY
 $SU(N)$ super Yang-Mills theory the supermultiplet structure
 gives the multiplier $N^2$ in front of QFT CA which should be 
 compared with
holographic one. The AdS$_3$/CFT$_2$ predicts the correspondent 
multiplier to be 
$N$. In AdS$_7$/CFT$_6$ correspondence where recently constructed 
(0,2) tensor multiplet plays the role of exact CFT the supermultiplet 
structure predicts the multiplier $N^3$ in front of QFT CA which should be 
compared with holographic one. It is very natural to expect then that
 AdS$_9$/CFT$_8$ correspondence exists (as well as respective d8 exact CFT) 
and the number of fields in CA appears with the factor $N^4$. As far as
 we know several attempts to construct d8 exact CFT are on the way currently.}
, the total 
anomaly should be given by
\be
\label{SSano2}
{\cal W}_{8} =N^4 \left( T_{\rm scalar} 
+T_{\rm spinor}\right)=- {7N^4 \over 6\left(2\pi\right)^4 l^8}\ .
\ee
By comparing (\ref{SSano2}) with (\ref{8danS}), we find
\be
\label{SSano3}
{1 \over \kappa^2}={216 N^4 \over 8867 
\left(2\pi\right)^4 l^7}\ ,
\ee
which might be useful to establish the proposal for AdS$_9$/CFT$_8$. 

Of course, the above relation gives only the indication (the numerical 
factor is definitely wrong) as we considered only scalar-spinor part of
non-conformal multiplet.
 On the same time it is known that 
for AdS$_7$/CFT$_6$ correspondence the tensor multiplet gives brane CFT while
 for d4 the gauge fields play the important role (super Yang-Mills theory).
As far as we know the rigorous proposal for d8 brane CFT does not exist yet.
However, it is evident that not only scalars and spinors but also
other fields will be part of d8 CFT. It would be extremely interesting to
 construct the candidate for such theory.
Then the above d8 holographic anomaly may be used to check the correctness 
of such proposal.

\ 

\noindent
{\bf Acknoweledgements} 

The work of S.O. was supported in part by Japan Society 
for the Promotion of Science and that of SDO was supported in 
part by CONACyT (CP, Ref.990356 and grant 28454E) and 
in part by RFBR. We are grateful to S.J. Gates for helpful discussion. 
S.O. would like to thank M. Fukuma for useful
discussions about the Hamilton-Jacobi equation.

\end{document}